\documentclass[pdflatex,sn-mathphys-num]{sn-jnl}
\usepackage{graphicx}%
\usepackage{multirow}%
\usepackage{amsmath,amssymb,amsfonts}%
\usepackage{amsthm}%
\usepackage{mathrsfs}%
\usepackage[title]{appendix}%
\usepackage{xcolor}%
\usepackage{textcomp}%
\usepackage{manyfoot}%
\usepackage{booktabs}%
\usepackage{algorithm}%
\usepackage{algorithmicx}%
\usepackage{algpseudocode}%
\usepackage{listings}%
\usepackage{caption}
\usepackage{subcaption} 
\theoremstyle{thmstyleone}%
\theoremstyle{thmstyletwo}%

\theoremstyle{thmstylethree}%

\begin{document}

\title[Article Title]{Regression modeling of multivariate precipitation extremes under regular variation}

\author[1]{\fnm{Rishikesh} \sur{Yadav}}\email{rishikesh@iitmandi.ac.in}
\author*[2]{\fnm{Arnab} \sur{Hazra}}\email{ahazra@iitk.ac.in}

\affil[1]{\orgdiv{School of Mathematical and Statistical Sciences}, \orgname{Indian Institute of Technology Mandi}, \orgaddress{\city{Mandi}, \postcode{175005}, \state{Himachal Pradesh}, \country{India}}}
\affil[2]{\orgdiv{Department of Mathematics and Statistics}, \orgname{Indian Institute of Technology Kanpur}, \orgaddress{\city{Kanpur}, \postcode{208016}, \state{Uttar Pradesh}, \country{India}}}


\abstract{
Motivated by the EVA2025 data challenge, where we participated as the team DesiBoys, we propose a regression strategy within the framework of regular variation to estimate the occurrences and intensities of high precipitation extremes derived from different climate runs of the CESM2 Large Ensemble Community Project (LENS2). Our approach first empirically estimates the target quantities at sub-asymptotic (lower threshold) levels and sets them as response variables within a simple regression framework arising from the theoretical expressions of joint regular variation. Although a seasonal pattern is evident in the data, the precipitation intensities do not exhibit any significant long-term trends across years. Besides, we can safely assume the data to be independent across different climate model runs, thereby simplifying the modeling framework. Once the regression parameters are estimated, we employ a standard prediction approach to infer precipitation levels at very high quantiles. We calculate the confidence intervals using a nonparametric block bootstrap procedure. While a likelihood-based inference grounded in multivariate extreme value theory may provide more accurate estimates and confidence intervals, it would involve a significantly higher computational burden. Our proposed simple and computationally straightforward two-stage approach provides reasonable estimates for the desired quantities, securing us a joint second position in the final rankings of the EVA2025 conference data challenge competition.
}
\maketitle
\section{Introduction}
\label{sec:introduction}
Extreme precipitation events are among the most common consequences of climate variability and change, with severe implications for water resource management, agriculture, and disaster preparedness. Extreme rainfall can cause ecological disturbances, landslides, floods, and infrastructure failures, resulting in high socioeconomic costs \citep{masson2021climate, Kundzewicz2014}. Climate research and hydrometeorology have thus made it a top priority to understand and model the statistical features and physical causes of precipitation extremes. It is often challenging to infer the underlying probability structure of extreme occurrences from rare observational records, as they are frequently local and uncommon. Therefore, rigorous statistical models that can extrapolate beyond observed data are currently in demand to estimate the likelihood and severity of heavy precipitation under different climatic conditions \citep{katz2010statistics}.

With the rise of ensemble climate modeling, researchers now have access to large datasets that capture internal variability and model uncertainty under different forcing scenarios. The CESM2 Large Ensemble Community Project (LENS2) \citep{Kay2015, rodgers2021ubiquity} provides a valuable resource, comprising several Earth System Simulation realizations designed to measure climate response variability. Analyzing extremes across multiple model runs allows us to assess not only marginal characteristics, such as return levels, but also inter-run dependencies and structural uncertainties. However, such analyses present statistical challenges: extremes are heavy-tailed, dependent, and frequently sparse in both space and time, which makes these analyses statistically challenging. In this regime, traditional Gaussian or regression-based approaches are inadequate, motivating the development of models grounded in extreme value theory (EVT). EVT provides the mathematical foundation for modeling rare and high-impact events \citep{deHaan2006, Coles2001}. In its simplest form, EVT describes the limiting distribution of maxima or threshold exceedances, which leads to, respectively, the Generalized Extreme Value (GEV) and Generalized Pareto Distribution (GPD) families \citep{Coles2001}. Researchers have widely utilized these models for analyzing precipitation and temperature extremes \citep{Kharin2013, Westra2014}; however, their adaptation to multivariate and spatial contexts poses significant challenges due to complex dependence structures. The notion of \emph{regular variation} serves as an essential tool for formalizing such dependencies \citep{Resnick2007, hult2006regular}, providing a way to examine joint tail behavior, define asymptotic dependence measures, and construct estimators for extremal coefficients that quantify the co-occurrence probabilities of rare events.

In environmental statistics, the modeling of joint or spatial extremes has made advances through various methodological approaches. Max-stable processes \citep{Schlather2002, padoan2010likelihood, davison2012statistical} define the asymptotic limits of normalized maxima and have emerged as a conventional tool for spatial modeling of extremes. It has already been effectively utilized for modeling extreme precipitation, wind speed, and temperature datasets. However, they are computationally intensive and show limited flexibility for extensive, high-dimensional datasets \citep{huser2022advances}. Copula-based methods offer an alternative by defining dependence structures independently from marginal distributions \citep{aas2009pair, Sibuya1960}. Still, the majority of conventional copulas (e.g., Gaussian or Student-$t$) poorly represent extremal dependence in the tails. The conditional extremes framework proposed by \cite{heffernan2004conditional} and further extended by \cite{wadsworth2012dependence} provides an alternative, flexible approach, allowing for the modeling of both asymptotic dependence and independence regimes via regression-type formulations.

There has been an increasing interest in regression methods for modeling tail dependence and other extreme functionals in the context of regular variation \citep{wadsworth2012dependence, engelke2020graphical}.  These models integrate the interpretability and scalability of regression techniques with the theoretical precision of tail limit outcomes.  In particular, linear or log-linear relationships often arise from the asymptotic properties of multivariate regularly varying vectors, which connect tail indices or scaling parameters across components.  These methods are ideal for large datasets, such as climate ensembles, where the objective is to estimate relationships among various sources, regions, or variables while preserving theoretical consistency.  Furthermore, one may utilize bootstrap-based resampling methodologies to evaluate estimation uncertainty in the context of temporal dependence \citep{politis1999subsampling, lahiri2013resampling}.

Motivated by the EVA2025 data challenge, we build upon this emerging line of research and propose a regression strategy under the framework of regular variation to estimate the occurrences and intensities of high precipitation extremes derived from multiple climate model runs of CESM2 LENS2. Our methodology begins with the empirical estimation of competition target quantities at sub-asymptotic thresholds, i.e., focusing on exceedances at moderate threshold levels. Treating these empirical quantities as response variables in a regression model formulated from the theoretical construction of joint regular variation enables the model to link empirical observations to asymptotic scaling laws, allowing for extrapolation to more extreme levels of precipitation intensity. A primary practical consideration in our modeling is the independence of extreme precipitation intensities across various climate model simulations.  Moreover, precipitation intensities measured in the unit system known as ``Leadbetter'' (as specified in the EVA2025 competition) exhibit minimal to no long-term trends and a weak serial correlation.  These features of the datasets simplify the joint modeling framework, which we utilize to fit regression equations relating the empirically estimated target quantities at different sub-asymptotic threshold levels as response variables, with the corresponding thresholds serving as covariates within a simple regression formulation.  We then use the predictions from these fitted regressions to calculate the competition target quantities at very high thresholds (beyond the observed data).

We use a nonparametric block bootstrap procedure to calculate the uncertainty in extrapolated estimates, treating years as blocks. The block bootstrap respects the dependence structure within each block (year) and provides robust confidence intervals for predicted target quantities. This combination of theory-based regression modeling and data-driven uncertainty quantification provides both interpretability and computational efficiency. Although our proposed approach is straightforward, it has proven effective in the EVA2025 competition. Our team, the DesiBoys, achieved a joint second position among all participating groups, demonstrating the practical utility of the framework for large-scale multivariate extreme data. Beyond the competition, this work makes significant contributions to the broader literature on extreme value modeling in several ways. First, it demonstrates how to apply regular variation-based regression methods in a real-world, high-dimensional climate-relevant context. Second, it gives us a precise and reproducible framework that strikes a balance between theoretical rigor and practical feasibility. Third, our method employs bootstrap-based inference to provide a flexible approach for measuring the uncertainty in predictions of rare events. These features make the proposed methodology useful for various applications in environmental and geophysical sciences that involve multivariate extremes.

The rest of this paper is organized as follows. Section~\ref{sec:EDA} briefly describes the CESM2 LENS2 dataset, provides exploratory analysis finding the key features of the data, and briefly recalls the EVA2025 challenge setup. Section~\ref{sec:modeling} presents the theoretical basis of regular variation and details the proposed regression modeling framework. Section~\ref{sec:results} reports main findings and uncertainty assessment. Finally, Section~\ref{sec:conclusion} summarizes the main contributions and outlines potential directions for future research.

\section{Data description and exploratory analysis}
\label{sec:EDA}
\subsection{Data description and task overview}
\label{subsec:data_details_task}
In this section, we briefly describe the data and problem statements tasked by the organizing committee,  and provide relevant exploratory analysis that motivated our modeling framework. 

The EVA 2025 Data Challenge focused on predicting extreme precipitation using simulated climate data from multiple runs. The dataset comes from the \href{https://www.cesm.ucar.edu/community-projects/lens2}{CESM2 Large Ensemble Community Project (LENS2)}, which provides multiple long-term climate model runs for 165 years (1850–2014) under historical forcing scenarios. For this data challenge, precipitation values (variable \texttt{PRECT}) were extracted over a $5 \times 5$ grid of neighboring locations. To anonymize and standardize the challenge, the exact locations were hidden, and the original precipitation units were transformed into an artificial unit called \texttt{Leadbetters}. We were provided with daily data from four randomly chosen ensembles, while the full ensembles consisted of 50 runs, 46 of which were withheld to evaluate the performance of the participating teams. Overall, the provided dataset consists of 165 years of daily precipitation data (transformed in \texttt{Leadbetters}) across 25 (5$\times$5) locations. Every year consists of 365 days, regardless of whether it is a leap year. From now on, throughout the manuscript, we will refer to precipitation intensities, which we mean as scaled precipitation amount in \texttt{Leadbetters}. Below are the three ultimate quantities of interest tasked by the organizing committee:
\begin{itemize}
    \item[\textbf{Task 1}:] Expected number of times all 25 locations exceed 1.7 Leadbetters,
    \item[\textbf{Task 2}:] Expected number of times at least 6 of the 25 sites exceed 5.7 Leadbetters,
   \item[\textbf{Task 3}:] Expected number of times at least 3 of the 25 sites exceed 5 Leadbetters for a run of at least two consecutive days.
\end{itemize}
Along with providing point estimates, the participating teams were asked to provide $95\%$ confidence intervals for all the above three questions.  
\subsection{Key features of the data}
\label{subsec:keydetailsdata}
To simplify the modeling assumptions, we conduct exploratory analysis, including long-term trend analysis, seasonality analysis, and temporal dependence analysis. 
\begin{figure}[ht]
    \centering
    \begin{subfigure}[t]{0.48\textwidth}
        \centering
        \includegraphics[height=0.65\linewidth]{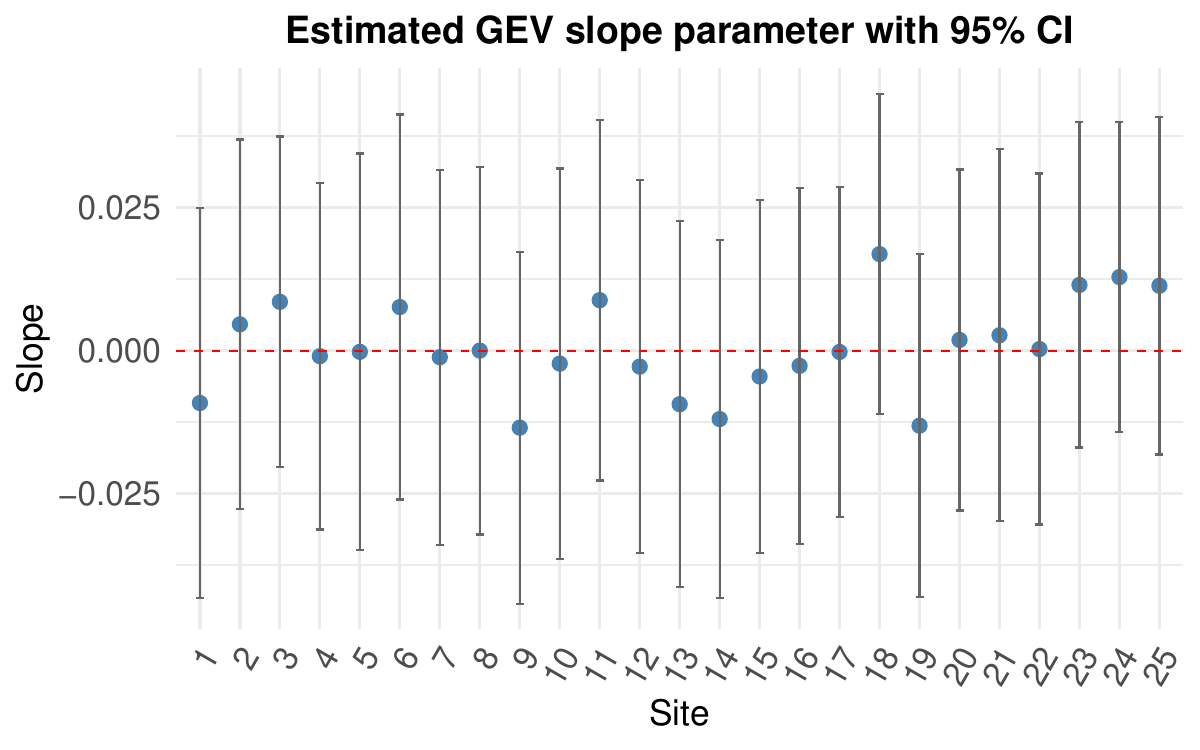}
    \end{subfigure}
        \hfill 
    \begin{subfigure}[t]{0.48\textwidth}
        \centering
        \includegraphics[height=0.65\linewidth]{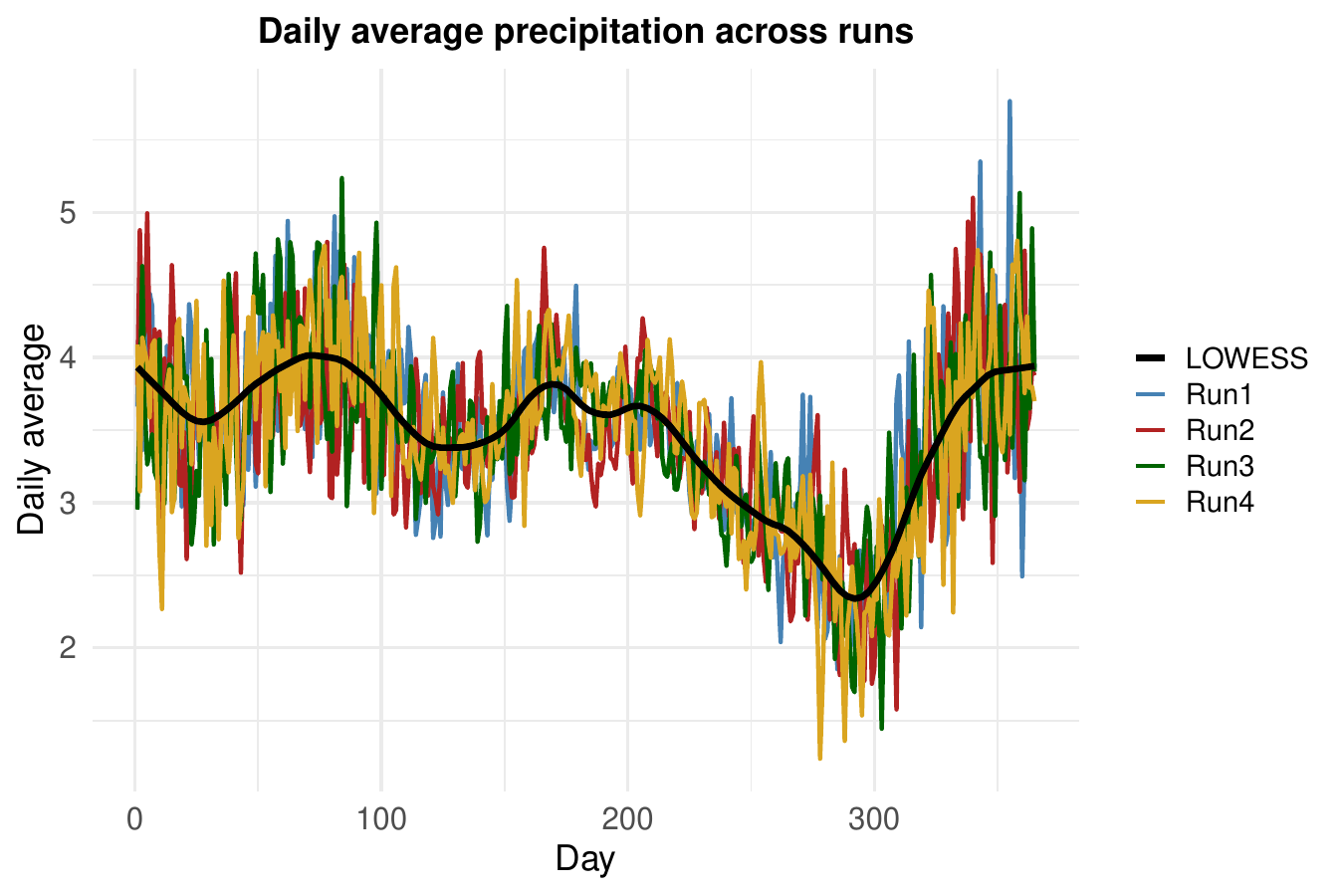}
    \end{subfigure}%
        \caption{\textit{Left:} Estimated slope components (blue dots) of the GEV location parameter fitted to yearly maxima, with 95\% confidence intervals (vertical bar). \textit{Right}: Daily average precipitation across four runs with LOWESS smoothing.}
         \label{fig:GEV_slope_and_seasonality}
\end{figure}

To examine temporal trends in the data, we fit a generalized linear model, site-wise, with GEV-distributed responses. Rescaled years, along with intercepts, are included as a linear predictor in the location parameter of the GEV distribution, while the scale and shape parameters are kept constant over time.  We fit the GEV model to the yearly maxima data while also including the yearly maxima data from four runs. The left panel of Figure \ref{fig:GEV_slope_and_seasonality} shows the site-wise estimated slope (coefficients of years) parameter and 95\% confidence intervals. From this plot, it is evident that there is generally no significant trend in the yearly maxima for any of the sites; this is a crucial assumption that enables us to simplify our modeling approaches; see Section \ref{sec:modeling} for further details. 

To check for seasonality in the data, we average the values across years and all sites for each day of the year, resulting in averaged daily data for the entire year. The right panel in Figure \ref{fig:GEV_slope_and_seasonality} displays the aggregated precipitation intensities for all four runs. From this plot, we can observe the clear presence of seasonality in the data. Furthermore, after removing the seasonality, we calculate the empirical temporal extremal dependence at the first lag for quantile levels of $0.975$, $0.98$, $0.985$, and $0.99$. For three out of four runs, the measure is calculated to be zero, indicating weak temporal extremal dependence. 


\begin{table}[ht]
\centering
\caption{
Summary of statistical tests for independence and stationarity for randomly selected sites. 
\textit{First row:} shows the $p$-values for run-test that checks for randomness in the sequence; a $p$-value $>$ 0.05 indicates the sequence is consistent with independence. 
\textit{Second row:} shows the $p$-values for the Augmented Dickey-Fuller (ADF) test that tests for stationarity; a very small $p$-value (typically $<$ 0.05) indicates the time series is stationary.
}
\begin{tabular}{c c c c c c c c c c c}
\hline
Site  & 3    & 5    & 6    & 10   & 11   & 14   & 15   & 18   & 19   & 23   \\
\hline
$\textrm{RunsTest}_{p}$  & 0.09 & 0.44 & 0.70 & 0.70 & 0.48 & 0.09 & 0.16 & 0.82 & 0.28 & 0.24 \\
$\textrm{ADF}_{p}$  & $<$0.01 & $<$0.01 & $<$0.01 & $<$0.01 & $<$0.01 & $<$0.01 & $<$0.01 & $<$0.01 & $<$0.01 & $<$0.01 \\
\hline
\end{tabular}
\label{tab:iid_tests}
\end{table}

We also evaluate whether the yearly maxima are independent and identically distributed (IID).  Since we assume the data across runs are IID, we include the annual maxima from each run, giving us $165\times4$ observations. Table~\ref{tab:iid_tests} displays the $p$-values for the run-test (test for independence) and the Augmented Dickey-Fuller test (test for stationarity) for 10 randomly chosen sites.  From this result, we can conclude that the yearly maxima data are IID across time, with a few sites exhibiting non-significant results at the Type-I error 0.05 but weak independence, as indicated by $p$-values between 0.05 and 0.1.

\section{Our proposed methodology}
\label{sec:modeling}
In this section, we present our methodology for estimating the three target quantities described in Section~\ref{sec:EDA}. Our methodological framework begins with the empirical estimation of target quantities at subasymptotic threshold levels (i.e., lower quantiles), as detailed in Section~\ref{subsec:empirical_est}. We then proceed to the theoretical estimation of the same quantities within the framework of regular variation in Section~\ref{subsec:RV_est}. Finally, in Section~\ref{subsec:estimation_reg}, we introduce a regression-based approach that links the empirical and theoretical estimates to evaluate the target quantities at higher threshold levels.

\subsection{Empirical estimation}
\label{subsec:empirical_est}
Let $Y_{i,r,t_1,t_2}$ denote the precipitation intensity (in Leadbetter) at site \(i=1,\dots,25\), run \(r=1,\dots, R(=4)\), year \(t_1 =1, \ldots,T_1 (=165)\) and day \(t_2=1,\dots,T_2 (=365)\). Then, we empirically calculate the three target quantities mentioned in Section \ref{subsec:data_details_task}. For fixed \((r,t_1,t_2)\), we write the site-wise order statistics (ascending) across \(i\) as
\begin{align}
\label{eq:orderstatitics}
Y_{(1),r,t_1,t_2}\le \cdots \le Y_{(25),r,t_1,t_2}.
\end{align}
Then, for Task 1, the indicator for the event of all 25 sites exceeding 1.7 Leadbetters is 
\begin{equation*}
I_{r,t_1,t_2} \;=\; \mathbb{I}\{Y_{(1),r,t_1,t_2}>1.7\},
\end{equation*}
where $\mathbb{I}\{ \cdot \}$ is the indicator function. Therefore, the total empirical count (over runs, years, and days) is
\begin{equation}
\widehat{E}_1 \;=\; \sum_{r=1}^{R}\sum_{t_1=1}^{T_1}\sum_{t_2=1}^{T_2} I_{r,t_1,t_2}.
\label{eq:task1_quant}
\end{equation}
Also, since the 20th order statistic (ascending) is the smallest among the top 6 sites,  the indicator for the event of at least 6 sites out of 25 exceeding 5.7 Leadbetters is 
\begin{equation*}
J_{r,t_1,t_2} \;=\; \mathbb{I}\{Y_{(20),r,t_1,t_2}>5.7\},
\end{equation*}
and the total empirical count (over years, runs, and days) for Task 2 is
\begin{equation}
\widehat{E}_2 \;=\; \sum_{r=1}^{R}\sum_{t_1=1}^{T_1}\sum_{t_2=1}^{T_2} J_{r,t_1,t_2}.
\label{eq:task2_quant}
\end{equation}
Additionally, since the 23rd order statistic (ascending) is the smallest among the top 3 sites, the indicator for the event of at least 3 sites out of 25 exceeding 5 Leadbetters is 
\begin{equation*}
K_{r,t_1,t_2} \;=\; \mathbb{I}\{Y_{(23),r,t_1,t_2}>5\},
\end{equation*}
and the total empirical count (over years, runs, and days) for Task 3 is
\begin{eqnarray}
\nonumber  \widehat{E}_3 &=& \sum_{r=1}^{R} \left \lbrace\sum_{t_1=1}^{T_1}\sum_{t_2=1}^{T_2 - 1} K_{r,t_1,t_2} K_{r,t_1,t_2+1} + \sum_{t_1=1}^{T_1 - 1} K_{r,t_1,T_2} K_{r,t_1+1,1} \right\rbrace \\
\nonumber && - \sum_{r=1}^{R} \left \lbrace \sum_{t_1=1}^{T_1}\sum_{t_2=1}^{T_2 - 2} K_{r,t_1,t_2} K_{r,t_1,t_2+1} K_{r,t_1,t_2+2} \right. \\
&& \left.~~~~~~~~ + \sum_{t_1=1}^{T_1 - 1} \left( K_{r,t_1,T_2-1} K_{r,t_1,T_2} K_{r,t_1+1,1} + K_{r,t_1,T_2} K_{r,t_1+1,1} K_{r,t_1+1,2} \right)\right\rbrace.
\label{eq:task3_quant}
\end{eqnarray}
Here, in Task 3, a run of three days counts as one run and is not counted as two runs of two days each. Hence, we adjust the empirical count as above.

\subsection{Estimation based on regular variation}
\label{subsec:RV_est}
Under the assumptions of no long-term trend across years and independence across runs, the expected counts under Task 1 may be written as 
\begin{align*}
    \mathbb{E}(\widehat{E}_1) = \sum_{t_1}^{T1}\sum_{r=1}^{R}\sum_{t_2=1}^{T2} \Pr\left(Y_{(1),r,t_1,t_2} > 1.7\right)= R T_1\sum_{t_2=1}^{T2} \Pr\left(Y_{(1),r,t_1,t_2} > 1.7\right).
\end{align*}
Furthermore, we assume that the random vector 
$\mathbf{Y}_{r,t_1,t_2} = (Y_{1,r,t_1,t_2},\dots,Y_{25,r,t_1,t_2})^\top$ 
is multivariate regularly varying \citep{basrak2002characterization, resnick2008extreme} with index $\alpha>0$. In other words, there exists a scaling function $b(t)\to\infty$ and a non-null Radon measure $\nu(\cdot)$ on $\mathbb{P} = [0,\infty]^{25}\setminus\{\mathbf{0}\}$ such that
\[
t\,\Pr\!\left(\frac{\mathbf{Y}_{r,t_1,t_2}}{b(t)} \in \cdot\right) \xrightarrow{v} \nu(\cdot),
\]
where ``$\xrightarrow{v}$'' denotes vague convergence of measures and  
the measure $\nu(\cdot)$ characterizes the tail dependence structure and is homogeneous of order $-\alpha <0$; i.e.,
\begin{align*}
\nu(c \mathcal{A}) = c^{-\alpha}\nu(\mathcal{A}), \quad \text{for all } c>0,\; \mathcal{A}\subset\mathbb{P}.
\end{align*}
For a large threshold $u$, the joint exceedance probability behaves as
\begin{align*}
\Pr\!\left(Y_{(1),r,t_1,t_2} > u\right)
= \Pr\!\left(\mathbf{Y}_{r,t_1,t_2} \in (u,\infty]^{25}\right)
\sim \nu\!\left((1,\infty]^{25}\right)\,u^{-\alpha}\ell(u),
\end{align*}
where $\ell(u)$ is a slowly varying function, and $\nu((1,\infty]^{25}) = C_{\min}$ measures the joint tail mass when all sites exceed level~1. As we assume that the marginal components $Y_{i,r,t_1,t_2}$ are identically distributed across $r$ and $t_1$ and exhibit either independence or a weak asymptotic dependence structure; we may treat $C_{\min}$ as a scaling constant summarizing the joint tail behavior and denote it by $C_{t_2}$ for convenience, corresponding to the $t_2$-th day of the year. Hence, under regular variation, the expected count becomes
\begin{equation*}
\mathbb{E}(\widehat{E}_1)
= R T_1 \sum_{t_2=1}^{T_2} 
\Pr\!\left(Y_{(1),r,t_1,t_2} > u\right)
\approx R T_1 \sum_{t_2=1}^{T_2} C_{t_2}\,u^{-\alpha_{t_2}}.
\end{equation*}
Straightforward algebra shows that $\lim_{u \to \infty}\sum_{t_2=1}^{T_2} C_{t_2}\,u^{-\alpha_{t_2}} / C_{t^*_2}\,u^{-\alpha_{t^*_2}} = 1$, where $t_2^* = \arg \min_{t_2 \in \{ 1,\ldots, T_2\}} \alpha_{t_2}$, assuming that $t_2^*$ is unique. Thus, rewriting $\alpha_{t^*_2}$ by $\alpha_1$ and $RT_1 C_{t^*_2}$ by $C_1$ for convenience, the probability further simplifies to
\begin{equation}
\mathbb{E}(\widehat{E}_1)
\approx C_1\,u^{-\alpha_1}. 
\label{eq:E1_model_based}
\end{equation}

The regular variation framework used for Task~1 extends naturally to Tasks~2 and~3 once the corresponding tail events are properly specified.
Recall that the target of Task~2 is the event that at least six of the $25$ sites exceed a high threshold $u$ on the same day.  
Using the ascending order statistics in \eqref{eq:orderstatitics},
this event may be written as $\{Y_{(20),r,t_1,t_2}>u\}$.  We define the corresponding tail set
\begin{align*}
\mathcal{A}_2
=
\Big\{\mathbf y\in[0,\infty]^{25}:\#\{i:y_i>1\}\ge 6\Big\}.
\end{align*}
Suppose the daily spatial vector $
\mathbf{Y}_{r,t_1,t_2}
=
\left(Y_{1,r,t_1,t_2},\dots,Y_{25,r,t_1,t_2}\right)^\top
$
is multivariate regularly varying with index $\alpha_2>0$ and tail measure $\nu_2$.  
Then, for large $u$,
\begin{align*}
\Pr\!\left(Y_{(20),r,t_1,t_2}>u\right)
=
\Pr\!\left(\mathbf{Y}_{r,t_1,t_2}\in u\,\mathcal{A}_2\right)
\sim
\nu_2(\mathcal{A}_2)\,u^{-\alpha_2}\,\ell_2(u),
\end{align*}
where $\ell_2(\cdot)$ is slowly varying. Under the assumptions of no long-term trend and identical tail behavior across runs and years, the expected empirical count satisfies
\begin{equation}
    \mathbb{E}(\widehat{E}_2)
=
R T_1 \sum_{t_2=1}^{T_2}
\Pr\!\left(Y_{(20),r,t_1,t_2}>u\right)
\approx
C_2\,u^{-\alpha_2},
~~
\textrm{where}~~C_2=\nu_2(\mathcal{A}_2).
\end{equation}
For Task 3, assuming a weak temporal extremal dependence across the daily observations, which is common in daily precipitation datasets (see Section \ref{subsec:keydetailsdata}),  the dominating term of the empirical count \eqref{eq:task3_quant} and its expectation are
\begin{eqnarray}
\nonumber \widehat{E}^*_3 &=& \sum_{r=1}^{R} \sum_{t_1=1}^{T_1}\sum_{t_2=1}^{T_2 - 1} K_{r,t_1,t_2} K_{r,t_1,t_2+1},\\
\mathbb{E}(\widehat{E}^*_3) &=& R T_1 \sum_{t_2=1}^{T_2 - 1} \textrm{Pr}(Y_{(23), r, t_1, t_2} > 5, Y_{(23), r, t_1, t_2+1} > 5). \nonumber
\end{eqnarray}
To reflect this structure, we define the two-day space--time vector
\begin{align*}
\mathbf{Z}_{r,t_1,t_2}
=
\big(\mathbf{Y}_{r,t_1,t_2}^\top,\,
      \mathbf{Y}_{r,t_1,t_2+1}^\top\big)^\top
\in [0,\infty]^{50},
\end{align*}
and the corresponding tail set
\begin{align*}
\mathcal{A}_3
=
\Big\{\mathbf z\in[0,\infty]^{50}:
\#\{i:z_i^{(1)}>1\}\ge3,\;
\#\{i:z_i^{(2)}>1\}\ge3
\Big\},
\end{align*}
where $z^{(1)}$ and $z^{(2)}$ denote the first- and second-day components, respectively. Assuming $\mathbf{Z}_{r,t_1,t_2}$ to be regularly varying with index $\alpha_3>0$ and space--time tail measure $\nu_3$, for large $u$, we have
\begin{align*}
\textrm{Pr}(Y_{(23), r, t_1, t_2} > u, Y_{(23), r, t_1, t_2+1} > u)
=
\Pr\!\left(\mathbf{Z}_{r,t_1,t_2}\in u\,\mathcal{A}_3\right)
\sim
\nu_3(\mathcal{A}_3)\,u^{-\alpha_3}\,\ell_3(u).
\end{align*}
Thus, under weak short-range temporal extremal dependence, the expected number of runs satisfies
\begin{equation}
  \mathbb{E}(\widehat{E}_3)
\approx
C_3\,u^{-\alpha_3},
\qquad
\textrm{where}~~C_3=\nu_3(\mathcal{A}_3).  
\end{equation}

\subsection{Estimation using regression approach}
\label{subsec:estimation_reg}
Our ultimate task is to estimate the scaling constants $C_1$, $C_2$, and $C_3$, as well as the shape parameters $\alpha_1$, $\alpha_2$, and $\alpha_3$, corresponding to three target quantities of interest based on the provided data. This task involves equating the empirical quantities from Section \ref{subsec:empirical_est} with those based on regular variation from Section \ref{subsec:RV_est} in a regression setup. Thresholds $u$ are selected to be moderate to sufficiently large, ensuring an adequate amount of data for empirical probability estimation and enabling the formulation of a regression model. Specifically, we choose different thresholds $u_l, l=1,\ldots,L$ at regular intervals and calculate the corresponding expected counts empirically; say $Z_l, l=1,\ldots,L$ denote the expected counts at threshold level $u_l$. Then, we may fit a simple linear regression of the form
\begin{align}
Z_l = C u_l^{-\alpha}, \qquad \text{or} \qquad \log(Z_l) = C'  + \alpha' \log(u_l), l=1,\ldots L, 
\label{eq:regression_equation}
\end{align}
to estimate the scaling constants $C$ and $\alpha$, where $C = \exp(C')$ and $\alpha = -\alpha'$. The above expressions are general one, and we may adopt it for any of the three target quantities. 

One of the requirements for the competition is to provide confidence intervals for the expected target counts for all three tasks.  We propose using a nonparametric block-bootstrap approach to calculate these intervals, treating years as blocks. Our choice is based on the fact that there is no trend over time, and the blocks (years) of data are independent.  We randomly sample the years with replacement to create 500 bootstrapped datasets.  Then, the 95\% confidence intervals are obtained by calculating the 2.5\% and 97.5\% quantiles of the estimated target quantities.

\section{Result}
\label{sec:results}
\begin{figure}[t!]
    \centering
    \includegraphics[width=\linewidth]{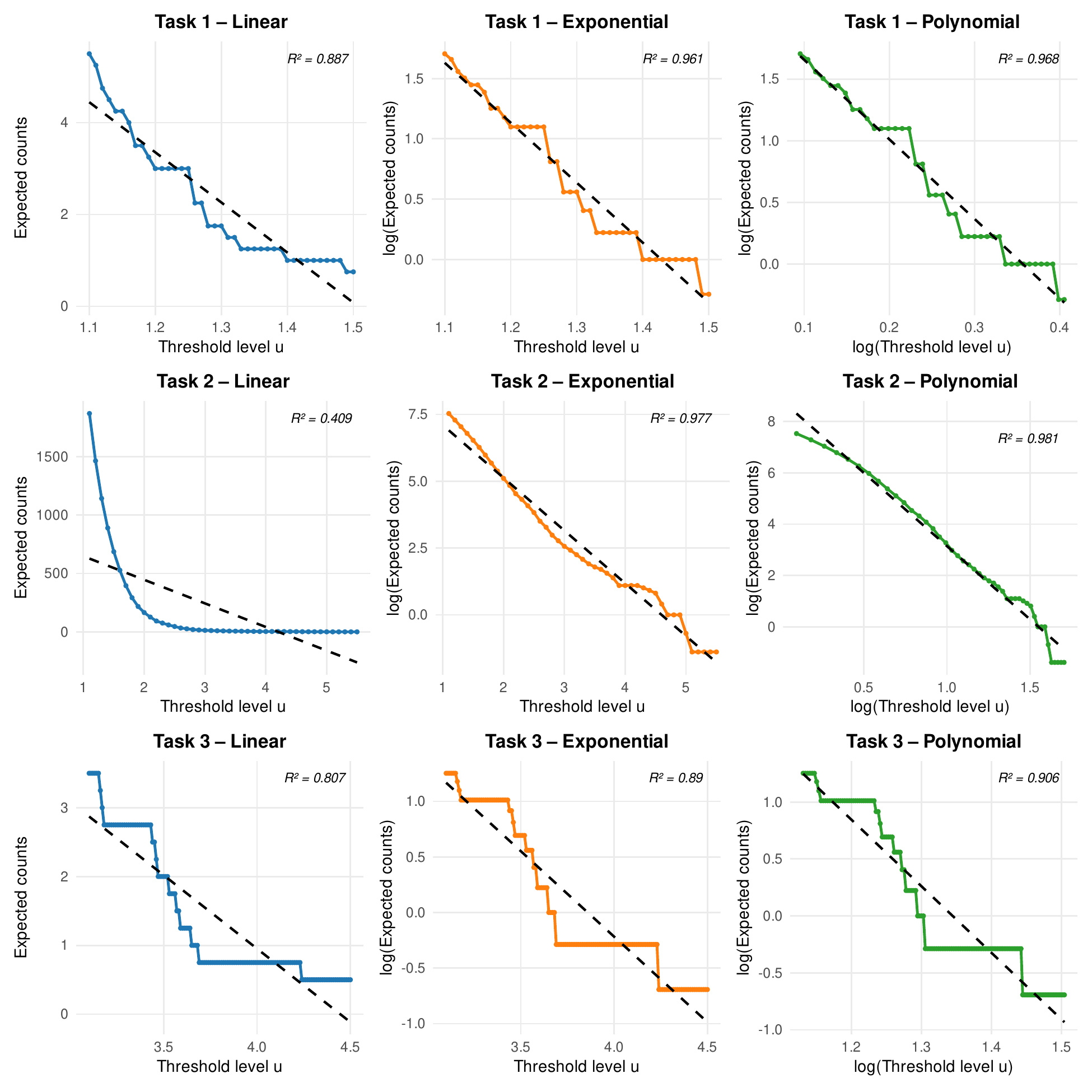}
    \caption{The relationship between thresholds, $u_l$, and empirical counts, $(Z_l)$, for Task~1 (first rows), Task~2 (second rows), and Task~3 (third row), shown across different transformations (column-wise). Corresponding $R^2$ values are also shown for respective regression fits to evaluate the goodness of fit.}
    \label{fig:Expl_p1}
\end{figure}
In this section, we provide details of the results submitted for the competition, along with the post-assessment after the competition results were published. For brevity, we do not discuss the results from the preliminary submission, although we followed similar approaches for both submissions.
\subsection{Details of submitted results}
\begin{figure}
    \centering
    \includegraphics[width=1\linewidth]{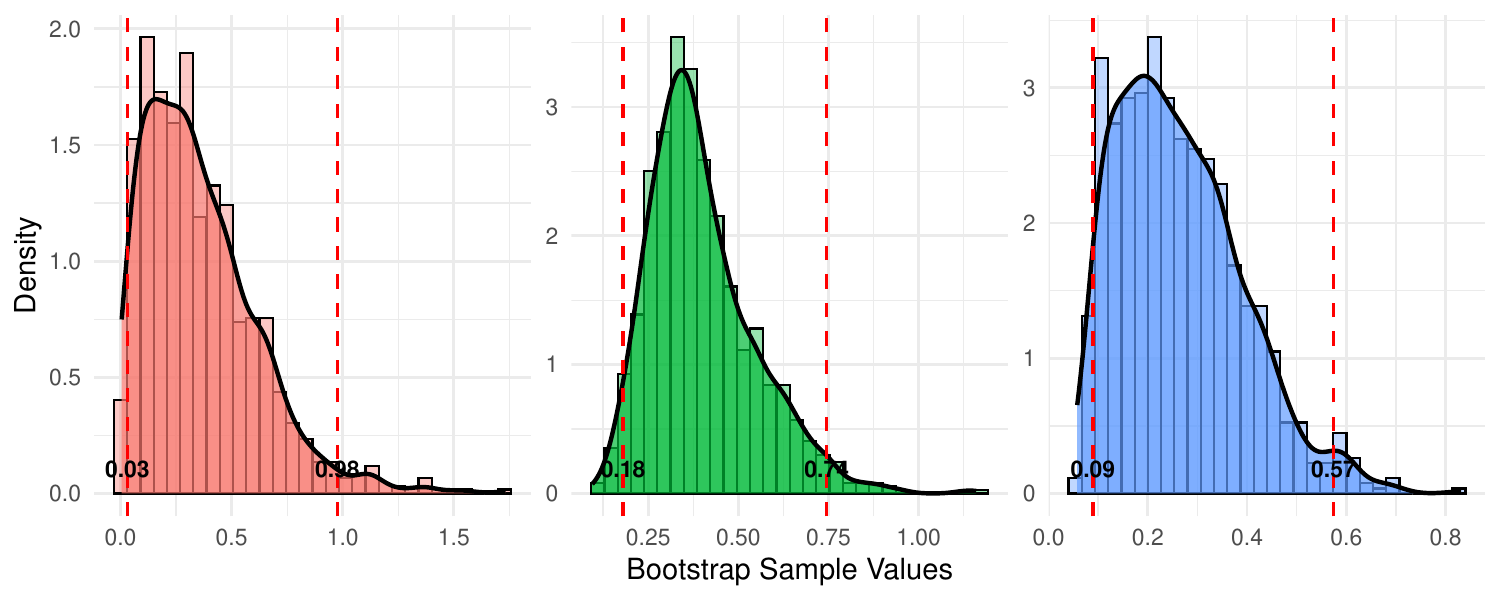}
    \caption{Histograms for the bootstrap samples for Task~1 (left), Task~2 (middle), and Task~3 (right) quantities.}
    \label{fig:bootstrap_CIs}
\end{figure}

For Task~1, we select the thresholds $u_l = 1.1, 1.11, 1.12, \ldots, 1.49, 1.50$, ensuring a sufficient amount of data exceeds these thresholds to derive the empirical estimates via Eqn. \eqref{eq:task1_quant}. Figure \ref{fig:Expl_p1} shows the expected empirical counts for Task~1 across several transformations, showcasing a clear linear relationship between $\log(u)$ and the logarithm of the expected counts (last column), also evident by the $R^2$ values, which gives close to $0.97$. This observation provides strong support for using the regression formula \eqref{eq:regression_equation}, resulting from the regular variational approach as discussed in Section \ref{subsec:estimation_reg}.  The estimates (sd) of the intercept parameters is $C_1' = 2.3 (0.05)$ and the slope parameter is $\alpha_1' = -6.44 (0.19)$ equivalent to $\alpha_1 = 6.44$ suggesting a heavy tail. Given these estimated parameters, we may obtain the estimates of Task~1 by setting $u=1.7$ in the  $C_1 u^{-\alpha_1} = \exp(2.3) \times 1.7^{-6.44} = 0.33$. The confidence intervals are obtained using nonparametric block bootstraps, which involve generating 500 bootstrap samples by blocking the entire dataset by year and subsequently reordering these years. The estimated 95\% confidence interval for the Task~1 quantity is $(0.03, 0.98)$; see also the left panel of Figure~\ref{fig:bootstrap_CIs} for the bootstrap distribution of Task~1 target quantities. 

Similarly, for the second task, we choose $u_l =1.1,1.2,\ldots,5.49, 5.5$, and fit a simple linear regression model of the form \eqref{eq:regression_equation} with the response variables as the empirical count given by \eqref{eq:task2_quant} calculated for each threshold $u_l$, which works as a predictor variable. The estimated parameters with SD shown in brackets are log-intercept $C_2' = 8.86 (0.14)$ and the shape $\alpha_2 = 5.71 (0.12)$. Based on this estimated parameter, the estimated Task~2 quantity is $0.34$. Confidence intervals (CIs) are obtained based on block bootstraps similar to Task~1 and the 95\% CIs are $(0.18, 0.75)$; see also the middle panel of Figure~\ref{fig:bootstrap_CIs} for the bootstrap distribution.  
 
In the third task, we adopt a similar approach by selecting thresholds $u_l = 1.10, 1.11, \ldots, 4.49, 4.50$ and fitting a simple linear regression model between the empirical counts~\eqref{eq:task3_quant} and the thresholds $u_l$. 
 The estimated log-intercept and slope parameters are $C_3' = 7.12\,(0.22)$ and $\alpha_3 = 5.28\,(0.16)$, respectively. 
 Using these estimates, we predict for the third task as \( \exp(7.12) \times 5^{-5.28} = 0.25. \)
 The 95\% confidence intervals are derived using the block bootstrap, as in Tasks 1 and 2, resulting in the interval $(0.09, 0.57)$; see also the right panel of Figure~\ref{fig:bootstrap_CIs} for the bootstrap distribution.

\subsection{Final performance and post-assessment}
The EVA2025 data competition organizers evaluated the performance of each team by establishing distinct criteria for both point and interval estimates. Point estimates were scored using the squared error, where the ``true'' estimates were obtained empirically by averaging the number of exceedances of the target event across all 50 runs (4 runs were provided for the competition, and another 46 were set aside for testing). The interval estimates were evaluated using the interval scores \cite{gneiting2007strictly}.
We performed reasonably well across all sub-competitions; specifically, we secured first position for point estimation in Task 3, while in Tasks 1 and 2, we achieved third and sixth place, respectively. However, our performance for the confidence intervals was comparatively on the lower side, as we ranked 7th, 5th, and 5th in Tasks 1, 2, and 3, respectively. Although block-bootstrap procedures are flexible and assumption-light, they may provide comparatively wider (narrower) intervals for estimators when the chosen block size is too large (small). In our application, we select annual blocks, assuming them to be identically distributed; however, a thorough investigation into a more appropriate block size choice could have improved our performance. Overall, in the combined performances of both point and interval estimation, our results were satisfactory, as we secured second place, tied with another team in the competition.  

\section{Discussion and future work}
\label{sec:conclusion} 
The proposed regression strategy, grounded in the theory of regular variation, provides a practical and theoretically coherent framework for modeling multivariate precipitation extremes.  The method connects empirically estimated sub-asymptotic quantities to asymptotic scaling laws through a regression formulation, thereby bridging the gap between purely empirical estimation and conventional extreme value theory.  We observe that extreme precipitation intensities show very little dependence between runs and no significant long-term trend. This observation suggests that large-scale model variability has a minimal effect on tail behavior.  This empirical finding has substantial implications for ensemble-based climate risk assessment, as it promotes the use of relatively straightforward dependence structures without compromising accuracy. The bootstrap-based estimation of confidence intervals provides a computationally efficient alternative to fully parametric likelihood-based methods, while ensuring robust uncertainty quantification.

Despite its strengths, several aspects of the present framework require further refinement. First, our regression formulation is based on a linear relationship derived from asymptotic regular variation; however, this may not fully account for nonlinearities or interactions between different spatial or temporal scales.  Adding more flexible function spaces, like additive models, Gaussian process priors, or penalized splines, could make predictions more accurate without making them harder to understand.  Second, although the independence assumption among climate model runs was empirically sufficient for this dataset, practical applications may necessitate explicit modeling of dependence across realizations, regions, or variables.  Expanding the methodology to incorporate such dependencies via hierarchical or graphical models \cite{engelke2020graphical} may indicate a promising future research direction. Furthermore, a more in-depth investigation into the impact of threshold selection on regression stability and extrapolation performance would enhance methodological robustness. In our current setup, we used a nonparametric bootstrap approach for the confidence intervals, which is often inappropriate as it tends to produce wider intervals for skewed distributions—precisely the case for this data competition; therefore, a more parametric approach would be necessary to better account for the uncertainty quantification of the extreme target quantities.

Moreover, the framework introduced here opens up multiple avenues for advancing the statistical modeling of environmental extremes across space and time.  One immediate extension may involve incorporating spatial and temporal covariates, such as topography and moisture flux, into the regression framework to enhance the understanding of variability in extremal behavior.  Combining this with spatial regular variation or max-stable representations could give us more information about how extremes spread from one area to another.  Finally, testing the proposed method on other climate variables, such as temperature or wind extremes, would demonstrate its effectiveness in a broader context and provide a unified perspective on assessing multivariate risk in the face of climate change.  All of these additions would make regular variation-based regression an even more helpful and accessible tool for looking at the tails of complex environmental systems.

\section*{Statements and Declarations}
\subsection*{Acknowledgment} We express gratitude to Dan Cooley (Colorado State University), Ben Shaby (Colorado State University), Jennifer Wadsworth (Lancaster University), and Emily Hector (University of Michigan) for organizing the EVA 2025 data competition. 

 \subsubsection*{Funding}
 Arnab Hazra acknowledges the financial support from the Start-up Research Grant (SRG/2023/000461) from the Science and Engineering Research Board (SERB), Government of India.

\subsubsection*{Conflict of interest}
The authors declare that they have no conflict of interest.

\bibliography{sn-bibliography.bib}

@article{gneiting2007strictly,
  title={Strictly proper scoring rules, prediction, and estimation},
  author={Gneiting, Tilmann and Raftery, Adrian E},
  journal={Journal of the American statistical Association},
  volume={102},
  number={477},
  pages={359--378},
  year={2007},
  publisher={Taylor \& Francis}
}

@book{resnick2008extreme,
  title={Extreme values, regular variation, and point processes},
  author={Resnick, Sidney I},
  volume={4},
  year={2008},
  publisher={{Springer Science and Business Media}},
  address={New York}
}

@article{basrak2002characterization,
  title={A characterization of multivariate regular variation},
  author={Basrak, Bojan and Davis, Richard A and Mikosch, Thomas},
  journal={The Annals of Applied Probability},
  volume={12},
  number={3},
  pages={908--920},
  year={2002},
  publisher={Institute of Mathematical Statistics}
}

@incollection{masson2021climate,
  title={Climate change 2021: the physical science basis},
  author={Masson-Delmotte, Val{\'e}rie and Zhai, Panmao and Pirani, Anna and Connors, Sarah L and P{\'e}an, Clotilde and Berger, Sophie and Caud, Nada and Chen, Y and Goldfarb, L and Gomis, MI and others},
  booktitle={Contribution of working group I to the sixth assessment report of the intergovernmental panel on climate change},
  volume={2},
  number={1},
  pages={2391},
  year={2021},
  publisher={Cambridge University Press},
  address={Geneva, Switzerland}
}

@article{Kundzewicz2014,
  title={Flood risk and climate change: global and regional perspectives},
  author={Kundzewicz, Zbigniew W and Kanae, Shinjiro and Seneviratne, Sonia I and Handmer, John and Nicholls, Neville and Peduzzi, Pascal and Mechler, Reinhard and Bouwer, Laurens M and Arnell, Nigel and Mach, Katharine and others},
  journal={Hydrological Sciences Journal},
  volume={59},
  number={1},
  pages={1--28},
  year={2014},
  publisher={Taylor \& Francis}
}

@article{katz2010statistics,
  title={Statistics of extremes in climate change},
  author={Katz, Richard W},
  journal={Climatic Change},
  volume={100},
  number={1},
  pages={71--76},
  year={2010},
  publisher={Springer}
}

@book{Coles2001,
  author    = {Stuart Coles},
  title     = {An Introduction to Statistical Modeling of Extreme Values},
  series    = {Springer Series in Statistics},
  publisher = {Springer},
  address = {London},
  year      = {2001}
}

@book{deHaan2006,
  title={Extreme value theory: an introduction},
  author={de Haan, Laurens and Ferreira, Ana},
  year={2006},
  publisher={Springer},
  address={New York}
}

@article{Kharin2013,
  title={{Changes in temperature and precipitation extremes in the CMIP5 ensemble}},
  author={Kharin, Viatcheslav V and Zwiers, Francis W and Zhang, Xuebin and Wehner, Michael},
  journal={Climatic Change},
  volume={119},
  number={2},
  pages={345--357},
  year={2013},
  publisher={Springer}
}

@article{Westra2014,
  title={Future changes to the intensity and frequency of short-duration extreme rainfall},
  author={Westra, Seth and Fowler, Hayley J and Evans, Jason P and Alexander, Lisa V and Berg, Peter and Johnson, Fiona and Kendon, Elizabeth J and Lenderink, Geert and Roberts, NM10},
  journal={Reviews of Geophysics},
  volume={52},
  number={3},
  pages={522--555},
  year={2014},
  publisher={Wiley Online Library}
}

@book{Resnick2007,
  title={Heavy-tail phenomena: probabilistic and statistical modeling},
  author={Resnick, Sidney I},
  year={2007},
  publisher={Springer},
  address={New York}
}

@article{hult2006regular,
  title={Regular variation for measures on metric spaces},
  author={Hult, Henrik and Lindskog, Filip},
  journal={Publications de l'Institut Math{\'e}matique},
  volume={80},
  number={94},
  pages={121--140},
  year={2006}
}

@article{Schlather2002,
  title={Models for stationary max-stable random fields},
  author={Schlather, Martin},
  journal={Extremes},
  volume={5},
  number={1},
  pages={33--44},
  year={2002},
  publisher={Springer}
}

@article{padoan2010likelihood,
  title={Likelihood-based inference for max-stable processes},
  author={Padoan, Simone A and Ribatet, Mathieu and Sisson, Scott A},
  journal={Journal of the American Statistical Association},
  volume={105},
  number={489},
  pages={263--277},
  year={2010},
  publisher={Taylor \& Francis}
}

@article{davison2012statistical,
author = {A. C. Davison and S. A. Padoan and M. Ribatet},
title = {{Statistical Modeling of Spatial Extremes}},
volume = {27},
journal = {Statistical Science},
number = {2},
publisher = {Institute of Mathematical Statistics},
pages = {161 -- 186},
year = {2012}
}

@article{aas2009pair,
  title={Pair-copula constructions of multiple dependence},
  author={Aas, Kjersti and Czado, Claudia and Frigessi, Arnoldo and Bakken, Henrik},
  journal={Insurance: Mathematics and Economics},
  volume={44},
  number={2},
  pages={182--198},
  year={2009},
  publisher={Elsevier}
}

@article{Sibuya1960,
  title={Bivariate extreme statistics},
  author={Sibuya, Masaaki and others},
  journal={Annals of the Institute of Statistical Mathematics},
  volume={11},
  number={2},
  pages={195--210},
  year={1960},
  publisher={Tokyo}
}

@article{heffernan2004conditional,
  title={A conditional approach for multivariate extreme values (with discussion)},
  author={Heffernan, Janet E and Tawn, Jonathan A},
  journal={Journal of the Royal Statistical Society Series B: Statistical Methodology},
  volume={66},
  number={3},
  pages={497--546},
  year={2004},
  publisher={Oxford University Press}
}

@article{wadsworth2012dependence,
  title={Dependence modelling for spatial extremes},
  author={Wadsworth, Jennifer L and Tawn, Jonathan A},
  journal={Biometrika},
  volume={99},
  number={2},
  pages={253--272},
  year={2012},
  publisher={JSTOR}
}

@article{engelke2020graphical,
  title={Graphical models for extremes},
  author={Engelke, Sebastian and Hitz, Adrien S},
  journal={Journal of the Royal Statistical Society Series B: Statistical Methodology},
  volume={82},
  number={4},
  pages={871--932},
  year={2020},
  publisher={Oxford University Press}
}

@book{lahiri2013resampling,
  title={Resampling methods for dependent data},
  author={Lahiri, Soumendra Nath},
  year={2013},
  publisher={Springer Science \& Business Media},
  address={New York}
}

@incollection{politis1999subsampling,
  title={{Subsampling in the IID Case}},
  author={Politis, Dimitris N and Romano, Joseph P and Wolf, Michael},
  booktitle={Subsampling},
  pages={39--64},
  year={1999},
  publisher={Springer},
  address={New York}
}

@article{Kay2015,
  title={The Community Earth System Model (CESM) large ensemble project: A community resource for studying climate change in the presence of internal climate variability},
  author={Kay, Jennifer E and Deser, Clara and Phillips, A and Mai, A and Hannay, Cecile and Strand, Gary and Arblaster, Julie Michelle and Bates, SC and Danabasoglu, Gokhan and Edwards, James and others},
  journal={Bulletin of the American Meteorological Society},
  volume={96},
  number={8},
  pages={1333--1349},
  year={2015}
}

@article{rodgers2021ubiquity,
  title={Ubiquity of human-induced changes in climate variability},
  author={Rodgers, Keith B and Lee, Sun-Seon and Rosenbloom, Nan and Timmermann, Axel and Danabasoglu, Gokhan and Deser, Clara and Edwards, Jim and Kim, Ji-Eun and Simpson, Isla R and Stein, Karl and others},
  journal={Earth System Dynamics},
  volume={12},
  number={4},
  pages={1393--1411},
  year={2021},
  publisher={Copernicus Publications G{\"o}ttingen, Germany}
}

@article{huser2022advances,
  title={Advances in statistical modeling of spatial extremes},
  author={Huser, Rapha{\"e}l and Wadsworth, Jennifer L},
  journal={Wiley Interdisciplinary Reviews: Computational Statistics},
  volume={14},
  number={1},
  pages={e1537},
  year={2022},
  publisher={Wiley Online Library}
}

\end{document}